\documentclass[twocolumn,showpacs,superscriptaddress,aps, prb]{revtex4-1}
\usepackage{amsmath}
\usepackage{amssymb}
\usepackage{graphicx}
\usepackage{sidecap}
\usepackage{float}
\usepackage[dvipdfm,colorlinks,linkcolor=green,urlcolor=blue, anchorcolor=blue,citecolor=green]{hyperref}
\usepackage{citesort}
\usepackage{xcolor}
\newcommand{\bq}{\begin{equation}}
\newcommand{\eq}{\end{equation}}
\newcommand{\bn}{\begin{eqnarray}}
\newcommand{\en}{\end{eqnarray}}

\begin{document}
%\input  epsf
%\draft
\title{Transient currents of a single molecular junction  with a vibrational mode}
\author {Guo-Hui Ding}
\affiliation{Key Laboratory of Artificial Structures and Quantum Control (Ministry of Education), Department of Physics and Astronomy,
  Shanghai Jiao Tong University, Shanghai, 200240, China}
\affiliation{Collaborative Innovation Center of Advanced Microstructures, Nanjing, China}

\author{Bo Xiong}
\affiliation{School of Physics and Astronomy, University of Minnesota, Minneapolis, Minnesota 55455, USA
}

\author{Bing Dong}
\affiliation{Key Laboratory of Artificial Structures and Quantum Control (Ministry of Education), Department of Physics and Astronomy,
  Shanghai Jiao Tong University, Shanghai, 200240, China}
\affiliation{Collaborative Innovation Center of Advanced Microstructures, Nanjing, China}
\date{\today }

\begin{abstract}
     By using a propagation scheme for current matrices and an auxiliary mode expansion method, we investigate the transient dynamics of a single molecular junction  coupled with  a vibrational mode. Our approach is  based on the Anderson-Holstein model and the dressed tunneling approximation for the electronic self-energy in the polaronic regime. The time-dependent currents after a sudden switching on the tunneling to  leads and an abrupt upward step bias pulse are calculated.   We show that the strong electron-phonon interaction greatly influences the nonlinear response properties of the system, and gives rise to interesting characteristics on the time traces of transient currents.

\end{abstract}

 \maketitle
% \narrowtext

\newpage

\section{Introduction}
The charge transport through nanoscale physical systems, including semiconductor quantum dots,\cite{vanderWiel2002}  single molecular junctions(SMJ), \cite{Galperin2007} and carbon nanotubes, \cite{Laird2015} etc., is currently under active investigations in the field of nanoelectronics.  During the electron tunneling processes,  the charging of a molecule or a nanotube quantum dot may cause deformations of its surrounding geometry and the emission or absorption of phonon(vibron) quanta,  inelastic electron tunneling effects due to the electron-phonon interaction (EPI) are ubiquitously observed. \cite{Leturcq2009,Park2000,Sapmaz2006}  It was found that the EPI can give rise to a variety of physically interesting phenomena in experiments on the dc characteristics of SMJs or the suspended carbon nanotube quantum dots, e.g., strong current suppression at low bias voltage termed Franck-Condon blockade \cite{Leturcq2009}, side peaks of differential conductance\cite{Park2000}, and negative differential conductances in the current-voltage characteristics\cite{Sapmaz2006}.

The dynamical properties of nanoscale systems under time-dependent external potentials are also of great interest and importance for their potential application in electronics devices. Some time-dependent potential related phenomena have been investigated for the systems in the linear response and the nonlinear regimes, for instance, the ac response and  current noise spectra,\cite{Ueda2011,Yu2014,Ding2014,Ding2013}  the quantum charge pumping\cite{DiCarlo2003,Leek2005,Croy2012} and the photon-assisted tunneling\cite{vanderWiel2002,Kohler2005}.  The theoretical investigation of time-dependent transport or transient  properties of nanoscale electronic systems is a challenging task, in particular for the systems with strong EPI or  strong Coulomb interaction. For a noninteracting quantum dot, a general theoretical framework for time-dependent transport properties has been given in the seminar work by Jauho {\it et al.} \cite{Jauho1994} and the switching effect of currents in response to sharp step- and also square-shaped voltage pulses has been analytically obtained by Maciejko {\it et al.}\cite{Maciejko2006}.  The transient current through a quantum dot with weak Coulomb interaction were studied by Schmidt {\it et al.}\cite{Schmidt2008} within the Anderson impurity model by using a mean field theory approximation and also the nonequilibrium Monte Carlo simulation method. The transient  and ac response properties of the Anderson impurity model in the Kondo regime were addressed by using the time-dependent noncrossing approximation.\cite{Nordlander1999,Nordlander2000,Plihal2000}   Recently, we have studied the time-dependent electron transport through a quantum dot with strong on-site Coulomb interaction \cite{Dong2015} using a second-order quantum rate equation and a propagation scheme proposed by Croy and Saalmann.\cite{Croy2009}

For the system with EPI, e.g., a SMJ or a carbon nanotube quantum dot coupled with vibrational mode, an archetype theoretical model is the Anderson-Holstein model. The transient effects of the Anderson-Holstein model in the weak EPI limit have been investigated using perturbation theory. \cite{Riwar2009} For a fast phonon mode ($\omega_0\gg \Gamma$) and in the case of sudden switching on the coupling to the leads, it was shown that the time-dependent total current exhibits plateau structures, which is attributed to the effect of electron shuttling due to the EPI. The transient dynamics of the Anderson-Holstein model in the strong EPI region is more complicated, a variety of theoretical methods were used to address this problem, e.g.,  nonequilibrium Green's function technique, \cite{Tahir2010} the polaron tunneling approximation, \cite{Albrecht2013} the diagrammatic Monte Carlo simulation approach, \cite{Muhlbacher2008,Schiro2009,Werner2009,Gull2011,Albrecht2012} multilayer multiconconfiguration time-dependent Hartree method\cite{Wang2009,Wang2011,Albrecht2012,Wilner2013,Wilner2014a,Wilner2014b}, etc.  Bistability of  steady state currents and the dot occupation number in the strong EPI case were shown for different initial preparations of the electronic or the phonon states of the system. \cite{Riwar2009,Albrecht2012,Wilner2013,Wilner2014a,Wilner2014b} However, a general consensus about the dynamics in the strong coupling region and the condition for bistability in the Anderson-Holstein model haven't been obtained.\cite{Klatt2015}

In this paper we  will study the transient dynamics of the Anderson-Holstein model in the strong EPI region by using the Lang-Firsov unitary transformation and a recently proposed approximation for the electron self-energy termed the dressed tunneling approximation (DTA). \cite{Dong2013,Souto2014} It was shown \cite{Souto2014} that the DTA  eliminates some
pathologies of the single-particle approximation (SPA) or the polaron tunneling approximation (PTA), and provides results for the current and the differential conductance in good agreement with that of the more elaborated methods such as the diagrammatic Monte Carlo simulation.\cite{Muhlbacher2008}
We consider the system in the polaronic regime with a fast phonon mode, and show that the time-dependent total current and also the displacement current  exhibit obvious signatures of EPI effect when the system is under a sudden switching on of the tunnel coupling to  leads or an upward step-shaped bias voltage. Therefore, the measuring of transient currents can give us
important information on the vibrational modes in the SMJs.

This paper is organized as follows: In Sec. II the model Hamiltonian and the self-energy terms in the presence of external time-dependent potentials are given. In Sec. III within the propagation scheme and the auxiliary-mode expansion method, we derive equations of motion for the dot occupation and the auxiliary current matrices. In Sec. IV numerical results and discussions on the transient currents are presented. In Sec. V  the summary and conclusion are given.

\section { Model Hamiltonian and the self-energy terms }
The Anderson-Holstein model can describe the electron transport between a molecular quantum dot(QD) and two metallic leads, where it is assumed that the molecular QD has only one energy level (with energy $\epsilon_d$) involved in the electron tunneling process, and is also coupled to a single vibrational mode (phonon) of the molecular with the frequency $\omega_0$ and the EPI strength $g_{ep}$. The Hamiltonian of this model is given as
\bn
H&=&\sum_{k\eta}\epsilon_{k\eta}(t)c^\dagger_{k\eta}c_{k\eta}+
 \epsilon_d(t) d^\dagger d +\omega_0 a^\dagger a + g_{ep} d^\dagger d(a^\dagger+a)
\nonumber\\
&&+\sum_{k\eta}\left [ \gamma_\eta(t) c^\dagger_{k\eta}d+{\rm H.c.}\right ]\;,
\en
where $\eta=L,R$ denotes the left and right leads,  $\epsilon_{k\eta}(t)=\epsilon_{k\eta}+\Delta_\eta(t)$ and $\epsilon_d(t) =\epsilon_d+\Delta_0(t)$,   with $\Delta_\eta(t)$ and $\Delta_0(t)$ being the time-dependent potentials applied to the lead $\eta$ and the dot region, respectively. $\gamma_\eta(t)$ describes the time-dependent tunneling-coupling matrix element between the QD and lead $\eta$, and takes the form of $\gamma_\eta(t)=\gamma_\eta u(t)$, with $u(t)$ being a time dependent function, e.g., in the case of sudden switching on the tunneling at $t=0$, $u(t)=\theta(t)$, with $\theta(t)$ denoting the Heaviside step function.  We can also define coupling strength between the lead and the QD as $\Gamma_\eta(\epsilon)=2\pi \sum_k |\gamma_\eta|^2\delta(\epsilon-\epsilon_k)$.   It should be noted that here we neglect the on-site Coulomb interaction in the molecular QD and consider a spinless electron model.  Applying the Lang-Firsov unitary transformation:\cite{Lang1963}  $\tilde H= e^S H e^{-S}$,
with $S=g d^\dagger d(a^\dagger-a)$ and the dimensionless parameter $g=g_{ep}/\omega_0$.  The transformed Hamiltonian becomes
\bn
\tilde H&=&\sum_{k\eta}\epsilon_{k\eta}(t)c^\dagger_{k\eta}c_{k\eta}+
 \bar\epsilon_d(t) d^\dagger d +\omega_0 a^\dagger a
\nonumber\\
&&+\sum_{k\eta}\left [ \gamma_\eta(t) c^\dagger_{k\eta}d X+\gamma^*_\eta(t) X^\dagger d^\dagger c_{k\eta} \right ]\;,
\en
where the phonon shift operator $X=e^{g(a-a^\dagger)}$,  $X^\dagger=e^{-g(a-a^\dagger)}$, and $\bar\epsilon_d(t)=\bar\epsilon_d+\Delta_0(t)$ with the renormalized energy level $\bar\epsilon_d=\epsilon_d-g_{ep}^2/\omega_0$. In the transformed Hamiltonian, the direct coupling between the electron and the phonon is eliminated, but the dot-lead tunneling amplitude is modified by the phonon shift operator $X$, which is responsible for the observation of the Franck-Condon steps in the dc current-voltage characteristics of the SMJs.\cite{Leturcq2009}

The electric current from the lead $\eta$ to the molecular dot $I_\eta(t)=-e\langle {\frac {d N_\eta} {dt}}\rangle$,  is given by
\bn
I_\eta(t)&=&{\frac {i e}{\hbar} }\sum_{k}\left [ \gamma_\eta(t) \langle c^\dagger_{k\eta}d X\rangle-\gamma^*_\eta(t) \langle X^\dagger d^\dagger c_{k\eta}\rangle
 \right ]\nonumber\\
 &=& {\frac {e} {\hbar}}[\Pi_\eta(t)+\Pi^*_\eta(t)]\;,
\en
where a current matrix $\Pi_\eta(t)\equiv i\sum_k\gamma_\eta (t)\langle c^\dagger_{k\eta}d X\rangle$ is introduced.
By using the equation of motion method, one can express the current matrix as an integration on the closed-time contour over the combination of the GFs of QD operator  and its corresponding self-energy as follows
\bn
\Pi_\eta(t)&=&\int_{-\infty}^t dt' [G^r_c(t,t')\Sigma^<_\eta(t',t)+G^<_c(t,t')\Sigma^a_\eta(t',t)]\nonumber\\
&=&\int_{-\infty}^t dt'[G^>_c(t,t')\Sigma^<_\eta(t',t)-G^<_c(t,t')\Sigma^>_\eta(t',t)]
\;.\nonumber\\
\en
Here $G_c(t,t')\equiv\langle T_C d(t)d^\dagger(t')\rangle$ denotes the GF of dot operator in the transformed Hamiltonian of Eq. (2), and  satisfies the Dyson equation on the closed-time contour
\bq
[i{\frac {\partial}{\partial t}}-\bar\epsilon_d(t)]G_c(t,t')=\delta(t,t')+\sum_\eta \int dt_1 \Sigma_\eta (t, t_1)G_c(t_1,t')\;.
\eq
 The self-energy term $\Sigma_\eta(t',t)$ is given by
\bq
\Sigma_\eta(t',t)=\sum_k\gamma^*_\eta(t')g_{k\eta}(t',t)\gamma_\eta(t)K(t,t')\;,
\eq
where $g_{k\eta}(t',t)$ is the GF of lead $\eta$ in the presence of the time-dependent potential $\Delta_\eta(t)$, and $K(t,t')$ is the propagator of the phonon shift operator, $K(t,t')\equiv\langle T_C X(t)X^\dagger(t')\rangle $. By making comparison with the noninteracting resonant tunneling model, one can see that here the self-energy due to tunneling into the leads is dressed by a polaronic cloud term.

In this work we consider the system with a fast phonon mode ($\omega_0\gg \Gamma$), and a strong energy dissipation of the vibration mode to a thermal bath, e.g., a substrate or a backgate. Then we can assume the phonons are always in equilibrium and have an equilibrium Bose distribution $n_B=1/(e^{\beta\omega_0}-1)$ at the temperature $T=1/\beta$.  The phonon shift operator GF $K(t,t')$ can be obtained by its equilibrium correlation function,\cite{Dong2013}  and its greater and lesser parts are given explicitly as
\bn
K^>(t,t')&=&\exp\{-g^2 [ n_B(1-e^{i\omega_0 (t-t')})\nonumber\\
        &+& (n_B+1) (1-e^{-i\omega_0 (t-t')})] \},
\en
and $K^<(t,t')=K^>(t',t)$. They can also expressed as sums of Fourier series

\bn
K^>(t,t')&=&\sum_{n=-\infty}^\infty w_n e^{-in\omega_0(t-t')}\;,\nonumber\\
K^<(t,t')&=&\sum_{n=-\infty}^\infty w_{-n} e^{-in\omega_0(t-t')}\;.
\en
The factor $w_n$ is the weighting factor describing the electronic tunneling involving absorption or emission of $n$ phonons.
\bq
w_n=e^{-g^2 (2n_B+1)}I_n(2g^2\sqrt{n_B(n_B+1)})e^{n\beta\omega_0/2}\;,
\eq
where $I_n(x)$ is the $n$th Bessel function of complex argument. Substitute Eq. (8) into the self-energy expression Eq. (6), we obtain
the greater and lesser parts of self-energy as follows
\begin{subequations}
\bn
\Sigma^>_\eta(t',t)&=&\sum_k \gamma^*_\eta(t')g^>_{k\eta}(t',t)\gamma_\eta(t)K^<(t,t')
\nonumber\\
&=&-i u(t') u(t)\sum_{n=-\infty}^\infty  w_n \int {\frac {d\epsilon} {2\pi}}[1-f_\eta(\epsilon)]
\nonumber\\&&\cdot \Gamma_\eta(\epsilon) e^{-i(\epsilon+n\omega_0)(t'-t)}e^{-i\int_{t}^{t'} d\tau \Delta_\eta(\tau)}\;,
\en
\bn
\Sigma^<_\eta(t',t)&=&\sum_k \gamma^*_\eta(t')g^<_{k\eta}(t',t)\gamma_\eta(t)K^>(t,t')
\nonumber\\
&=&i u(t') u(t)\sum_{n=-\infty}^\infty  w_{-n} \int {\frac {d\epsilon} {2\pi}}f_\eta(\epsilon)
\nonumber\\&&\cdot \Gamma_\eta(\epsilon) e^{-i(\epsilon+n\omega_0)(t'-t)}e^{-i\int_{t}^{t'} d\tau \Delta_\eta(\tau)}\;.
\en
\end{subequations}

Here $f_\eta(\epsilon)=1/[e^{\beta(\epsilon-\mu_\eta)}+1]$ is the Fermi distribution function,  with  $\mu_\eta$ being the chemical potential of lead $\eta$.

 \section{Auxiliary-mode expansion and the propagation scheme}
 In the above section we show that an energy integral over Fermi function is involved in the calculation of the lesser or greater self-energy. It will be a heavy burden on computation if multiple integral is encountered in the study of transient dynamics.   At finite temperature, a well-known method for performing the energy integral is to use the reside theorem and represent the integral by the Matsubara summation,\cite{Mahan} but this method is suffered from slow convergence especially at low temperature. Some highly accurate and more efficient methods have been developed in the literatures. In this work, we will use the auxiliary-mode expansion introduced by Croy and Saalman.\cite{Croy2009} The Fermi function is expanded in a sum over simple poles
 \bq
 f_\eta(\epsilon)\approx {\frac {1} {2}}-{\frac {1} {\beta}}\sum_{p=1}^{N_P}\left ({\frac {1} {\epsilon-\chi_{\eta p}^+}}+{\frac {1} {\epsilon-\chi_{\eta p}^-}} \right )\;,
 \eq
 where $\chi_{\eta p}^\pm=\mu_\eta\pm i\chi_p/\beta$ (with $\chi_p>0$). Thus all poles $\chi_{\eta p}^+$ ($\chi_{\eta p}^-$) are in the upper (lower) complex plane (details for how to calculate these pole parameters can be found in the appendix of Ref. [22]). $N_P$ is the number of pole pairs. Therefore, we will replace the Fermi function in Eq. (10) by the auxiliary-mode expansion above.

 Next, in order to take into account finite-bandwidth effects, we take the energy dependence of the linewidth function $\Gamma_\eta(\epsilon)$ to be Lorentzian
 \bq
 \Gamma_\eta(\epsilon)={\frac {\Gamma^0_\eta W^2} {\epsilon^2+W^2}}={\frac {\Gamma^+_\eta} {\epsilon-iW}}+{\frac {\Gamma^-_\eta} {\epsilon+iW}}\;,
 \eq
 where $\Gamma^0_\eta$ is the linewidth constant and $W$ is the bandwidth. It indicates that this linewidth function has two simple poles at $\epsilon=\pm i W$ with residues $\Gamma_\eta^{\pm}=\mp  {i\Gamma^0_\eta W}/2$. Then Eqs. (11) and (12) are plugged into the definition of the self-energies in Eq. (10), and  the energy integral is evaluated by the contour integration method. For $t>t'$, we close the contour in the upper half plane and pick up the poles to obtain

\begin{subequations}
 \bn
&&\Sigma^>_\eta(t', t)=u(t') u(t)\sum_n w_n \bigg \{ \Gamma^+_\eta[1-f^P_\eta(iW)] \nonumber\\
  &&\cdot e^{-i(n\omega_0+iW)(t'-t)}
  +\sum_{p=1}^{N_F} \Gamma_\eta(\chi^+_{\eta p})e^{-i(\chi^+_{\eta p}+n\omega_0)(t'-t)} \bigg \}\nonumber\\
  &&\cdot  e^{-i\int_t^{t'} d\tau \Delta_\eta(\tau)}\;,
\en
\bn
 && \Sigma^<_\eta(t', t)=-u(t') u(t)\sum_n w_{-n} \bigg \{ \Gamma^+_\eta f^P_\eta(iW) \nonumber\\
  &&\cdot e^{-i(n\omega_0+iW)(t'-t)}
  -\sum_{p=1}^{N_F} \Gamma_\eta(\chi^+_{\eta p})e^{-i(\chi^+_{\eta p}+n\omega_0)(t'-t)} \bigg \}\nonumber\\
  &&\cdot  e^{-i\int_t^{t'} d\tau \Delta_\eta(\tau)}\;,
 \en
\end{subequations}
where $f^P_\eta(iW)$ denotes that the Fermi function at the pole $iW$ calculated by the expansion given in Eq. (11). To simplify the notation,
one can introduce the coefficients and exponents as follows
\begin{subequations}
\bq
\Gamma^{>,\pm}_{\eta n\lambda}=\bigg \{\pm w_n \Gamma^{\pm}_\eta[1-f^P_\eta(\pm iW)], \pm {\frac {1} {\beta}} w_n\Gamma_\eta(\chi^{\pm}_{\eta p})\bigg \}\;,
\eq
\bq
\Gamma^{<,\pm}_{\eta n\lambda}=\bigg \{\mp w_{-n} \Gamma^{\pm}_\eta f^P_\eta(\pm iW), \pm {\frac {1} {\beta}} w_{-n}\Gamma_\eta(\chi^{\pm}_{\eta p})\bigg \}\;,
\eq
\bq
\chi^{\pm}_{\eta n \lambda}=\bigg \{ \pm i W+n\omega_0, \chi^{\pm}_{\eta p}+n\omega_0  \bigg \}\;.
\eq
\end{subequations}
Then the self-energies can be written in a compact form
\bq
\Sigma^{\gtrless}_\eta(t', t)=u(t)\sum_{n\lambda} \Sigma^{\gtrless}_{\eta n \lambda}(t',t)\;,
\eq
in which for $t>t'$,
\bq
\Sigma^{\gtrless}_{\eta n \lambda}(t',t)=u(t')\Gamma^{\gtrless,+}_{\eta n \lambda }e^{-i\chi^+_{\eta n\lambda}(t'-t)}e^{-i\int_{t}^{t'} d\tau \Delta_\eta(\tau)}\;,
\eq
and for $t<t'$, we have
\bq
\Sigma^{\gtrless}_{\eta n \lambda}(t',t)=u(t')\Gamma^{\gtrless,-}_{\eta n \lambda }e^{-i\chi^-_{\eta n\lambda}(t'-t)}e^{-i\int_{t}^{t'} d\tau \Delta_\eta(\tau)}\;.
\eq
The current matrix $\Pi_\eta(t)$ in Eq. (4) can be written in terms of  auxiliary-mode expansion current matrices
$\Pi_{\eta n \lambda}(t)$
\bq
\Pi_\eta(t)=u(t) \sum_{n\lambda} \Pi_{\eta n \lambda}(t)\;,
\eq
with
\bq
\Pi_{\eta n \lambda}(t)=\int_{-\infty}^t dt'[G^>_c(t,t')\Sigma^<_{\eta n \lambda}(t',t)-G^<_c(t,t')\Sigma^>_{\eta n \lambda}(t',t)]\;.
\\
\eq
By differential the above equation with respect to the time variable $t$, and follow the same procedure as in Ref. [22],  we obtain the equation of motion for the auxiliary current matrices
\bn
i {\frac  {\partial} {\partial t}}\Pi_{\eta n \lambda}(t)&=& u(t) \bigg [\Gamma^{<,+}_{\eta n\lambda}+ n_d(t)(\Gamma^{>,+}_{\eta n\lambda}-\Gamma^{<,+}_{\eta n\lambda})\bigg ]
\nonumber\\
&+& \bigg [\bar\epsilon_d+\Delta_0(t)-(\chi^+_{\eta n\lambda}+\Delta_\eta(t))\bigg ]\Pi_{\eta n \lambda}(t)
\nonumber\\
&+& u(t) \sum_{\eta' n' \lambda'} \Omega_{\eta n\lambda, \eta' n' \lambda'}(t)\;,
\en
where a new quantity $\Omega_{\eta n\lambda, \eta' n' \lambda'}(t)$ is introduced, which satisfies  the following equation of motion
\bn
&&{\frac {\partial}{\partial t}}\Omega_{\eta n\lambda, \eta' n' \lambda'}(t)=u(t)\bigg [\Pi_{\eta n \lambda}(t)(\Gamma^{>,-}_{\eta' n'\lambda'}
-\Gamma^{<,-}_{\eta' n'\lambda'}) \nonumber\\
&&\hspace*{1cm}+(\Gamma^{>,+}_{\eta n\lambda}-\Gamma^{<,+}_{\eta n\lambda})\Pi^*_{\eta' n' \lambda'}(t)\bigg ]
+i\bigg [(\chi^+_{\eta n\lambda}+\Delta_\eta(t)) \nonumber\\
&&\hspace*{1cm}-(\chi^-_{\eta' n'\lambda'}+\Delta_{\eta'}(t))\bigg ]\Omega_{\eta n\lambda, \eta' n' \lambda'}(t)\;.
\en

From the current conservation condition in the QD region, it is easy to see that equation of motion for the QD  occupation number $n_d(t)$ is given
by
\bn
{\frac {\partial}{\partial t}} n_d(t)&=&\sum_\eta[\Pi_\eta(t)+\Pi_\eta^*(t)]
\nonumber\\
&=&u(t)\sum_{\eta n \lambda}[\Pi_{\eta n \lambda}(t)+\Pi^*_{\eta n \lambda}(t)]\;.
\en
The Eqs. (20), (21) and (22) form a closed set of first order differential equations for current matrices in this auxiliary-mode expansion method, and give the basis of 
a propagation scheme for the transient dynamics of this EPI system.  

\begin{figure}[htp]
\begin{center}
\includegraphics[width=\columnwidth, height=3.5in ]{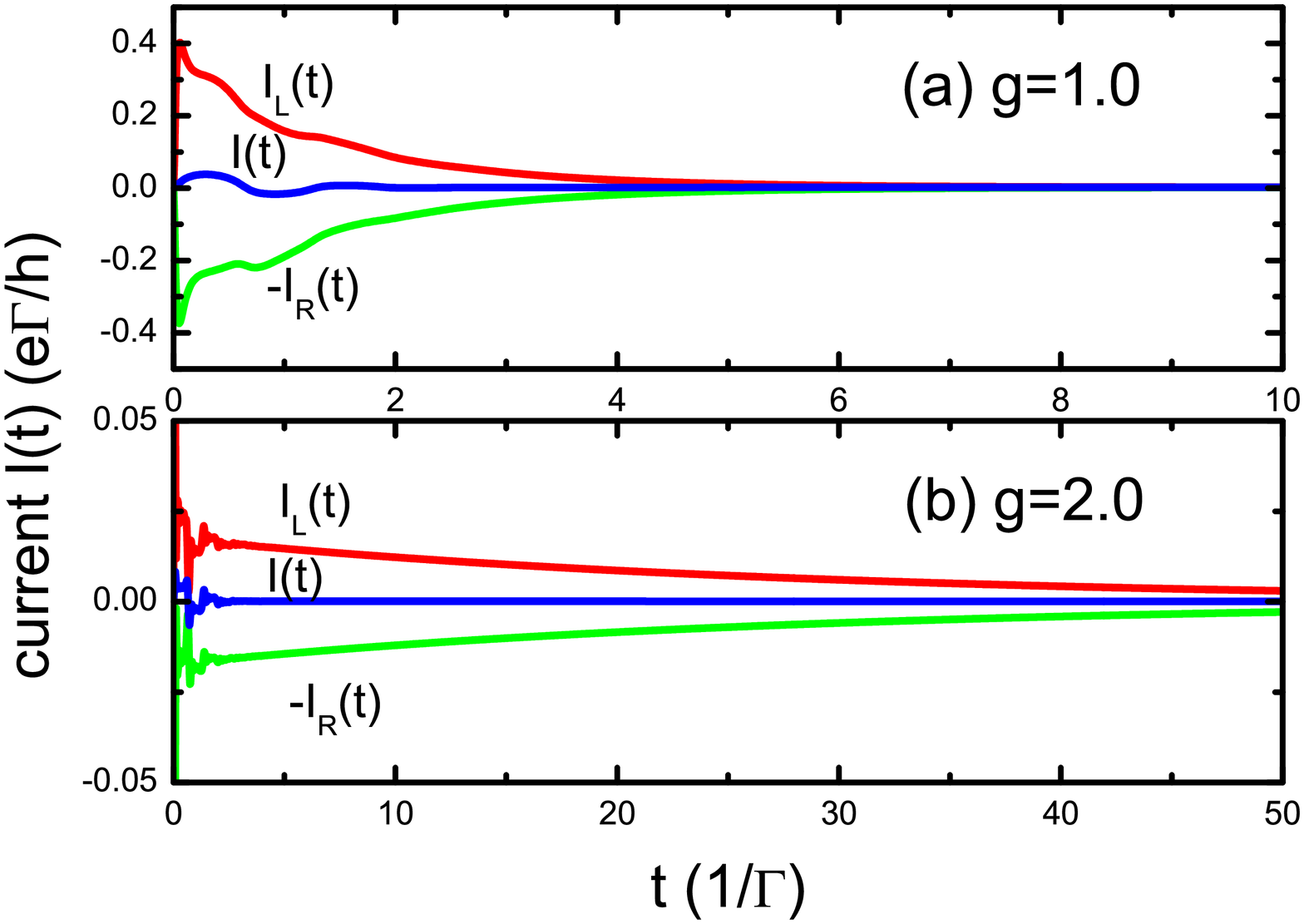}
\caption{Left, right and total current $I_L(t)$, $I_R(t)$ and $I(t)$  as a function of time at  different electron-phonon coupling constants: $g=1.0$ (a); $2.0$ (b), respectively. The energy parameters used are in units of $\Gamma$, with the bandwidth of leads $W=50$, $\Gamma_L^0=\Gamma_R^0=\Gamma=1.0$,  the bias voltage $\Delta\mu=2.0$, the frequency of vibration mode $\omega_0=10.0$, and the dot level $\bar\epsilon_d=-5.0$. The quantum dot is assumed initially in the empty state.}
\end{center}
\end{figure}

\section {Results and discussions}
 By solving the above set of partial differential equations with the finite difference method, we can obtain the numerical results of the transient current for various initial conditions.   For simplicity, we consider the system with symmetric tunnelling coupling to the leads, with $\Gamma_L^0=\Gamma_R^0=1.0$,  and  $\Gamma=(\Gamma_L^0+\Gamma_R^0)/2$ is taken as the unit of energy. The phonon frequency is assumed as $\omega_0=10.0$,  and the Fermi level of the leads at equilibrium give the reference point of energy, $\mu_L=\mu_R=0$.  In our numerical calculation, the system is taken as at a finite temperature $T=0.5$. The number of poles $N_P=80$ is used for the expansion of Fermi distribution function, which ensures the convergence of the numerical results. We study the transient properties of the current in this system, which can be divided into two terms: the total current $I(t)=[I_L(t)-I_R(t)]/2$, which measures the charge transported through the system,  and the displacement current $I_{disp}(t)=I_L(t)+I_R(t)$, which reflects the change of occupation number in the dot region $I_{disp}(t)={\dot n}_d(t)$.\cite{Riwar2009}

\begin{figure}[htp]
\begin{center}
\includegraphics[width=\columnwidth, height=3.0in ]{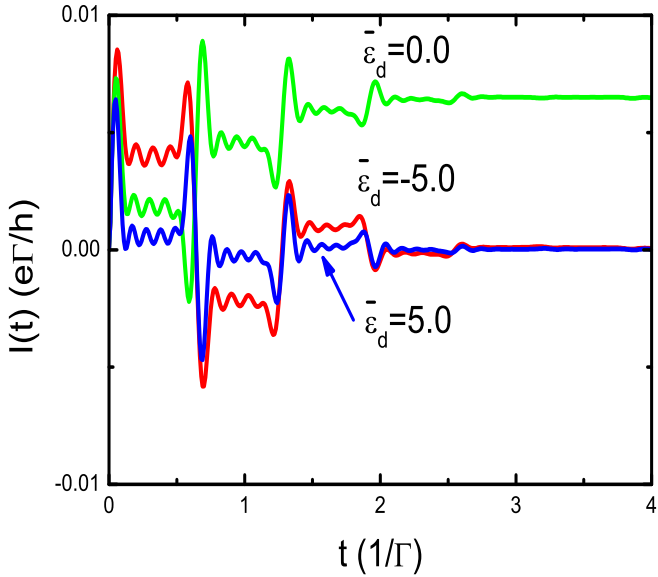}
\caption{Time-dependent total current  $I(t)$ for initially empty molecular quantum dot.  The electron-phonon coupling constants $g=2.0$, the bias voltage $\Delta\mu=2.0$. The other parameters used are the same as that in figure 1.}
\end{center}
\end{figure}

\begin{figure}[htp]
\begin{center}
\includegraphics[width=\columnwidth, height=3.0in ]{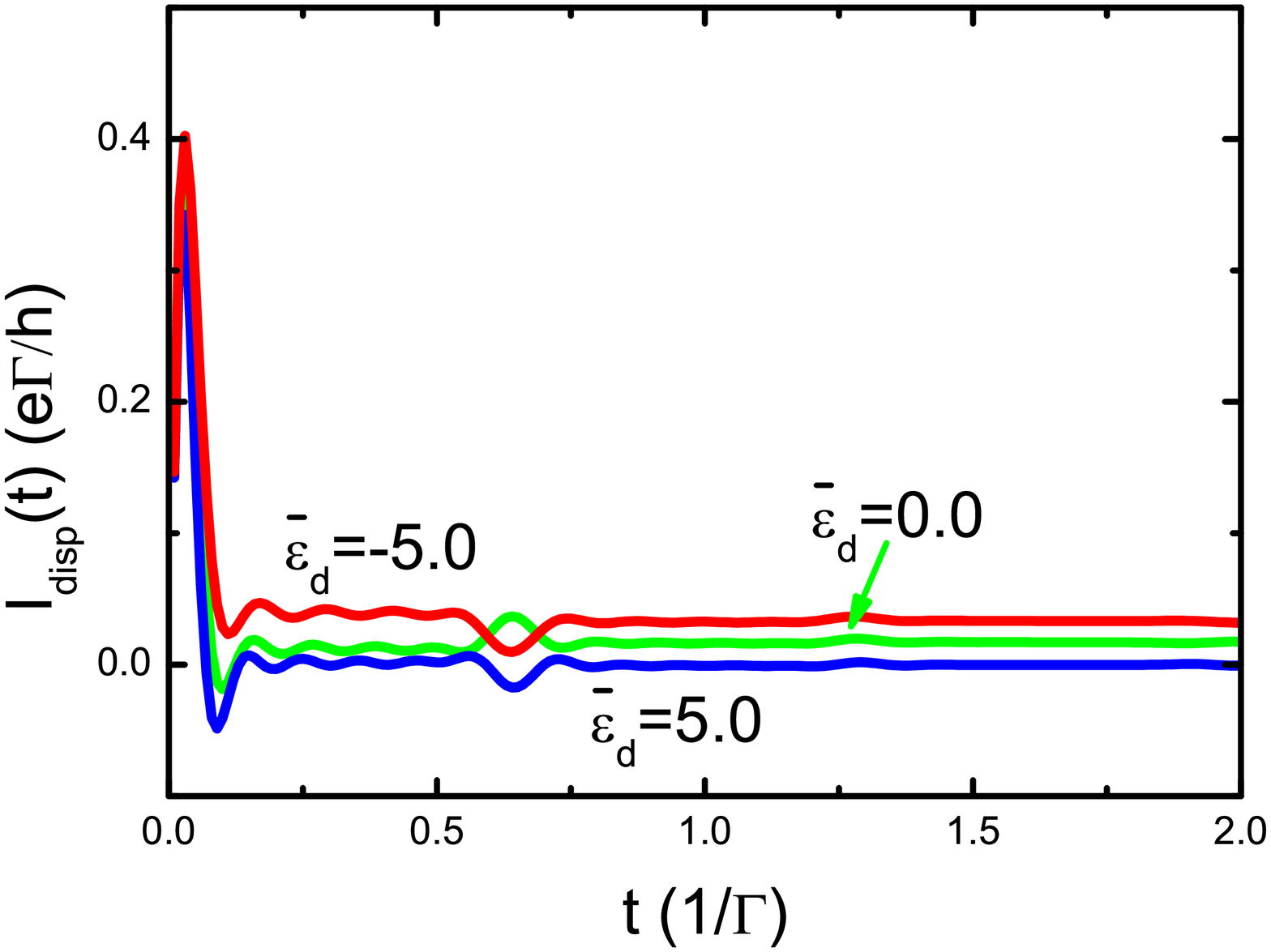}
\caption{The displacement current  $I_{disp}(t)$ for initially empty molecular quantum dot.  The electron-phonon coupling constants $g=2.0$, the bias voltage $\Delta\mu=2.0$.}
\end{center}
\end{figure}

 \subsection{ Sudden switching on of the coupling to leads}
We first consider transient current behaviors when the coupling of the quantum dot to leads is suddenly switched on at an initial time ($t_0=0$). In this case the time-dependent function $u(t)=\theta(t)$, and we will assume the quantum dot initially to be empty. In Fig. 1 the current $I_L$ and $I_R$ from the left and the right leads and also the total current $I(t)$  are plotted for the system with a bias voltage $\Delta\mu=2.0$. Because the finite band-width (we take $W=50$) of leads has been taken into account in our calculation, the current $I_L$ and $I_R$ are zero at the initial time $t=0$, then increase to the maximal values at a very short time-scale of $t\approx 1/W$. It is well known that in the infinitely wide band approximation, an unphysical current jump to a finite value immediately at the initial time is obtained. Therefore, the finite bandwidth approach adopted in this work gives a more accurate description of the transient current behavior.  The time evolution of currents in Fig. 1 shows that the system will reach to a steady state in the long time limit, however a much longer time is needed for the system with the strong EPI. One can attribute this phenomenon to the strong renormalization of the tunneling coupling between the leads and the quantum dot because of the EPI.  For weak EPI case ($g=1.0$) shown in Fig. 1(a), the time-dependence of  currents is relatively smooth.  However, for the strong EPI case ($g=2.0$) in Fig. 1(b),  more complex structures of current evolution are clearly indicated in the short time region.

In Fig. 2 we examine the total current in the short time region for the systems with strong EPI in more details. Three different dot level positions are considered for the SMJ under a fixed bias voltage ($\Delta\mu=2.0$). As the dot level $\bar\epsilon_d=0$, which is located in the transport window ($\mu_R<\bar \epsilon_d<\mu_L$), the steady total current in the long time limit has a relatively high value. On the contrary, the steady total current is very small when the dot level $\bar\epsilon_d$ is far below or above the Fermi levels of the leads. It shows that in the short time region the total current exhibits Fano lineshape in the time domain with  peak and  dip structures at the time $t\approx 2\pi n/\omega_0$, where $n$ is an integer. Similar current oscillating behaviors were also found previously by the time-dependent APTA and the diagrammatic Monte-Carlo simulation.\cite{Albrecht2013}, and were attributed to the electron shuttling due to the EPI.  This result indicates that experimental measuring of  the transient total current through SMJs can provide useful information about the vibration modes of the molecule between the leads.

 The displacement current $I_{disp}(t)$ which reflects the charge accumulation process in the dot is plotted in Fig. 3.  The displacement currents for three different dot levels all exhibit a sharp current peak initially, because the dot to leads coupling is suddenly switched on. For the dot level well above the Fermi level, negative value of $I_{disp}$ is found immediately after the initial current peak, and indicates that transient charge depletion processes in the center region can also exist in this sudden switching on case. At the time $t\approx 2\pi n/\omega_0$, there are also some peak or dip structures at the displacement current, which manifest the EPI effect.

\subsection{ Currents driven by an upward step pulse }
 Now, we consider another situation with the quantum dot initially coupled with left and right leads, and the system is in an equilibrium state with zero bias voltage.
Then a finite bias applied symmetrically between leads ($\mu_L=-\mu_R=\Delta\mu/2$) is suddenly turned on at the time $t_0=0$, and a transient current is driven through the quantum dot.

\begin{figure}[htp]
\begin{center}
\includegraphics[width=\columnwidth, height=3.5in ]{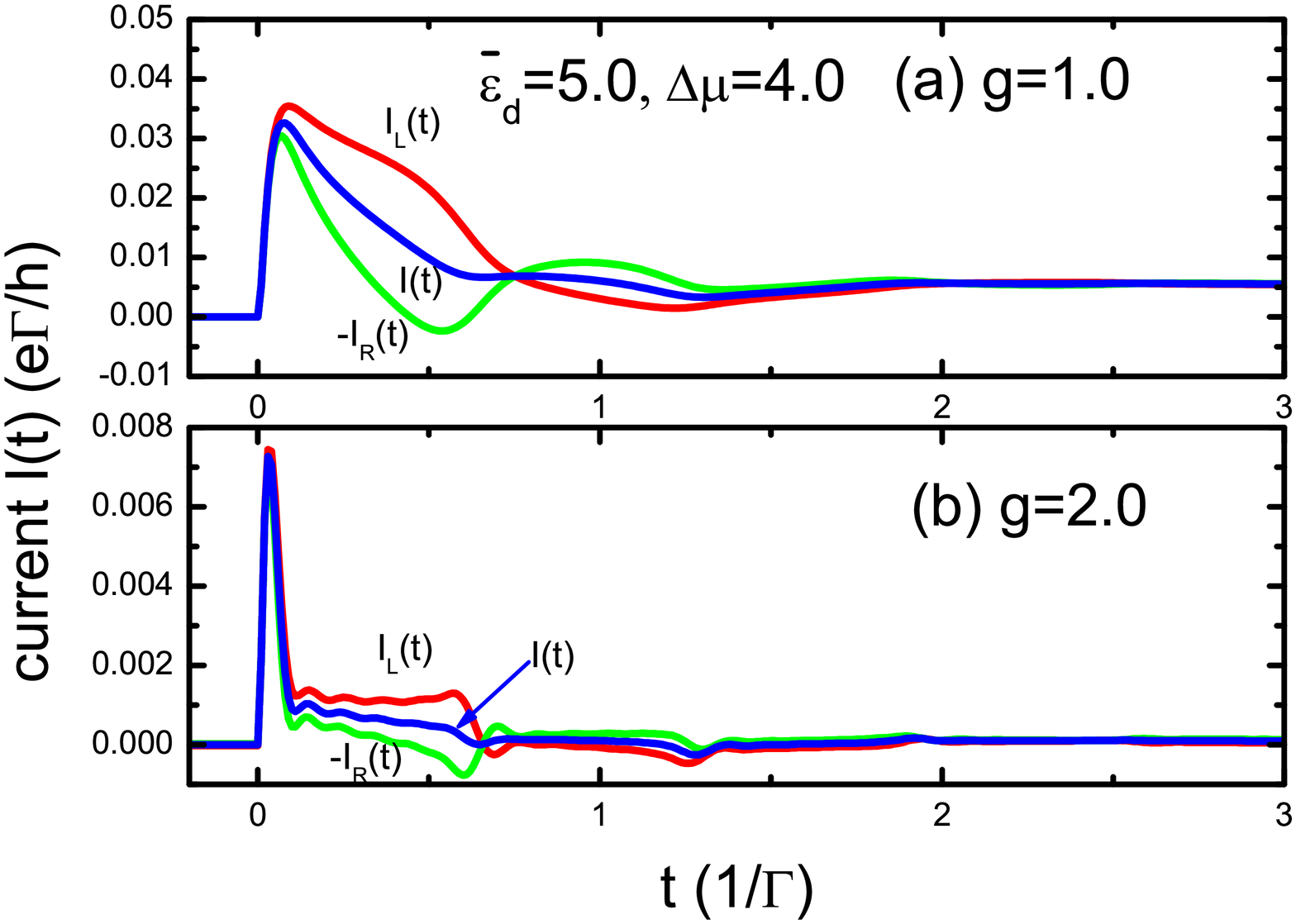}
\caption{Left, right and total current $I_L(t)$, $I_R(t)$ and $I(t)$ after an upward step pulse. The dot level $\bar\epsilon_d=5.0 $, and the electron-phonon coupling constants are: (a) $g=1.0$; (b) $g=2.0$ ,
respectively.  The bias voltage for the pulse $\mu_L=-\mu_R=2.0$.}
\end{center}
\end{figure}

In Fig. 4 the time-dependent left, right and  total current after an upward step pulse are plotted.  Both in Fig. 4(a) and (b)  sharp rising of current amplitude after  turning on the bias voltage are observed. For weak EPI case in Fig. 4(a),  all the left, right and total currents have one broad peak initially,  then there are some smooth current oscillations as the system evolves to the steady state. The magnitude of the left current $I_L(t)$ for electron tunneling into the quantum dot is not equal to that of the tunneling out current  $-I_R(t)$, which indicates that there are charge accumulating and depletion processes in the centre region. For the system with strong EPI ($g=2$), the first peak for the current becomes quite sharp, and the current amplitude also decreases drastically compared with the weak EPI case, which can be attributed to the strong renormalization of the tunneling coupling strength $\Gamma$ by EPI. In the time evolution of the currents, some periodic structures with a time-period of $2\pi/\omega_0$ are clearly observed, and it manifests the EPI effect in the system.

Next we study the total current under different step voltage pulses in the strong EPI case. Fig. 5(a) and (b) correspond to the systems with  different dot levels, respectively.  When the dot level $\bar \epsilon_d$ is  well above the Fermi energy ($E_F=0$) as shown in Fig. 5(a), significant increasing of the steady currents in the long-time limit is observed with increasing the bias voltage, whereas for  the dot level located around the Fermi energy shown in Fig. 5(b),  the steady current values are almost independent of the bias voltages. It is easy to understand this behavior since in the later case the dot level is already located in the transport window even for a small bias voltage.
Sharp initial current peaks are observed in both cases of dot level locations. Broad current peaks following the sharp peaks are exhibited for the system with $\bar\epsilon_d=0$, indicating a relatively larger current flow in this system after switching on the bias voltage.  As the bias voltage increases,  peak or dip structures of the total current arising from EPI interaction at the time $t=2 n\pi/\omega_0$  become more significant.

\begin{figure}[htp]
\begin{center}
\includegraphics[width=\columnwidth, height=3.5in ]{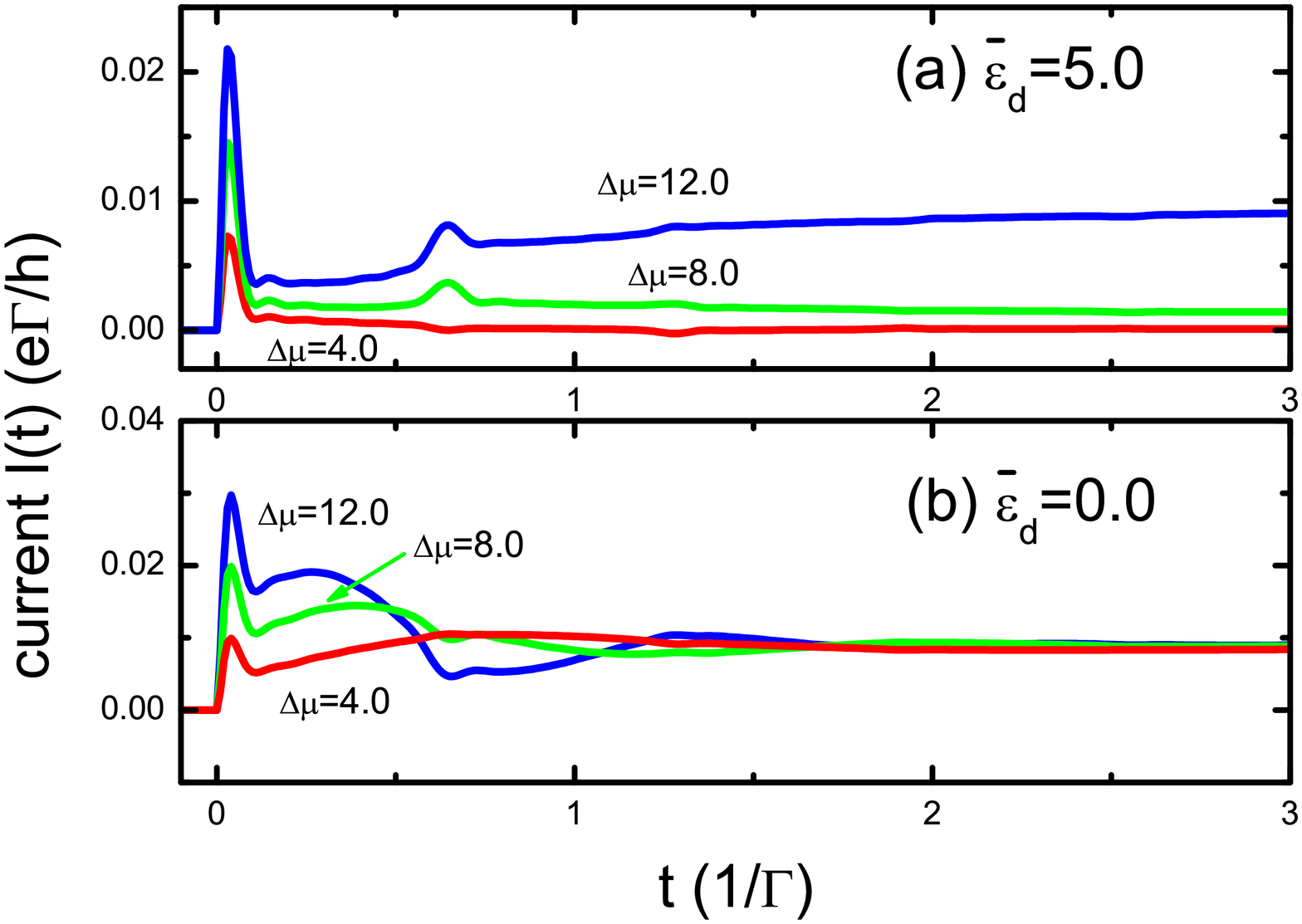}
\caption{The time-dependent total current  $I(t)$ for the quantum dot system under different upward step bias voltages. The dot level are: (a) $\bar\epsilon_d=5.0 $; (b) $\bar\epsilon_d=0.0$, respectively. The EPI constants is taken as $g=2.0$.}
\end{center}
\end{figure}

\section {Conclusion}
 We have investigated  transient effects in the electron tunneling through a SMJ coupled with a vibrational mode by using the nonequilibrium GF method and a propagation scheme for current matrices. We mainly focus on the system with the phonon mode frequency $\omega_0\gg \Gamma$ and  in the strong EPI region. The total current when the tunneling to the leads is suddenly switched on gives  evident EPI effects, such as the emergence of  plateau structures with a time period of $2\pi/\omega_0$, which reveals the frequency of vibrational mode in the molecular junction.  The long time for the system to reach a steady state represents a strong renormalization of the tunneling coupling between leads and the quantum dot by the EPI. As for the system under a upward step bias pulse, sharp initial current peaks are observed, but the oscillation of the total current is significantly suppressed in the strong EPI case.  EPI effects are also noticeable in the time trace of the total current by changing the bias voltage or tuning the energy level of the quantum dot. One may expect that the main features of transient dynamics of the SMJ shown in this work can be testified experimentally using ultrafast spectroscopy techniques.

\begin{acknowledgments}
{This work was supported by Projects of the National Basic Research Program of China (973 Program) under Grant No. 2011CB925603,
the National Natural Science Foundation of China (Grant Nos.91121021 and 11074166), and Shanghai Natural Science Foundation (Grant No. 12ZR1413300).}
\end{acknowledgments}

\bibliography{refsmoljunt}

\end{document}